\documentclass[journal]{IEEEtran}
\usepackage{tensor}
\usepackage{graphicx}
\usepackage{subcaption}
\usepackage{tabularx}
\usepackage[cmex10]{amsmath}
\usepackage{amsthm}
\usepackage{amssymb}
\usepackage{bm}
\usepackage{algorithm}
\usepackage{algorithmic}
\usepackage{cite}
\usepackage{setspace}
\usepackage{stfloats}
\usepackage{enumerate}
\usepackage{cases}
\usepackage{multirow}
\usepackage{mathrsfs}
\usepackage{array}
\usepackage{color}
\usepackage{verbatim}
\usepackage{extpfeil}
\usepackage{hyperref}
\usepackage{booktabs}
\usepackage{multirow}
\usepackage{caption}
\usepackage{graphicx}
\usepackage{tabularx}
\usepackage{epstopdf}
\usepackage{textcomp}
\usepackage{epsfig,epsf,color,balance,cite}
\usepackage{amssymb}
\usepackage{amsthm}
\usepackage{graphicx}
\usepackage{array,color}
\usepackage{amsmath}
\usepackage{graphicx}
\usepackage{tabularx}
\usepackage{epsfig,epsf,color,balance,cite}
\usepackage{algorithmic}
\usepackage{algorithm}
\usepackage{bm}
\usepackage{caption}
\usepackage{textcomp}
\usepackage{color}
\usepackage{multirow}
\usepackage{cite}
\usepackage{enumerate}
\usepackage{cases}
\usepackage{color}
\usepackage{epstopdf}
\usepackage{textcomp}
\usepackage{url}

\def\BibTeX{{\rm B\kern-.05em{\sc i\kern-.025em b}\kern-.08em
    T\kern-.1667em\lower.7ex\hbox{E}\kern-.125emX}}
\begin{document}

	\makeatletter
	\newcommand{\rmnum}[1]{\romannumeral #1}
	\newcommand{\Rmnum}[1]{\expandafter \@slowromancap \romannumeral #1@}
	\makeatother
	
	\title{Employing High-Dimensional RIS Information for RIS-aided Localization Systems}
	
	\author{  Tuo Wu, Cunhua Pan,  Kangda Zhi,  Hong Ren,   Maged Elkashlan, and Jiangzhou Wang, \emph{Fellow, IEEE}, \\Chau Yuen, \emph{Fellow, IEEE}
\thanks{\emph{(Corresponding author: Cunhua Pan.)}}
\thanks{T. Wu, K. Zhi and M. Elkashlan are with the School of Electronic Engineering and Computer Science at Queen
Mary University of London, London E1 4NS, U.K. (Email:\{tuo.wu,k.zhi, maged.elkashlan\}@qmul.ac.uk). C. Pan, H. Ren are with the National Mobile Communications Research Laboratory, Southeast University, Nanjing 210096, China. (e-mail: \{cpan, hren \}@seu.edu.cn).  J.
			Wang is with the School of Engineering, University of Kent, UK. (e-mail: J.Z.Wang@kent.ac.uk). C. Yuen is with the School of Electrical and Electronic Engineering, Nanyang Technological University, 639798, Singapore (e-mail: chau.yuen@ntu.edu.sg).  }
}
	
	\markboth{}
	{}
	\maketitle

	\begin{abstract}
		Reconfigurable intelligent surface (RIS)-aided localization systems have attracted extensive research attention due to their accuracy enhancement capabilities. However, most studies primarily utilized the base stations (BS) received signal, i.e., BS information, for localization algorithm design, neglecting the potential of RIS received signal, i.e., RIS information.  Compared with BS information, RIS information offers  higher dimension and richer feature set, thereby significantly improving the ability to extract positions of the mobile users (MUs). Addressing this oversight, this paper explores the algorithm design based on the high-dimensional RIS information. Specifically, we first propose a RIS information reconstruction (RIS-IR) algorithm to reconstruct the high-dimensional  RIS information from the low-dimensional BS information. The proposed RIS-IR algorithm comprises a data processing module for preprocessing BS information, a convolution neural network (CNN) module for feature extraction, and an output module for outputting the reconstructed RIS information. Then, we propose a  transfer learning based fingerprint (TFBF) algorithm that employs the reconstructed high-dimensional RIS information for MU localization. This involves adapting a pre-trained DenseNet-121 model to map the reconstructed RIS signal to the MU's three-dimensional (3D) position. Empirical results affirm that the localization performance is significantly influenced by the high-dimensional RIS information and maintains robustness against unoptimized phase shifts.
	\end{abstract}

	\IEEEpeerreviewmaketitle

\section{Introduction}
The advent of sixth-generation (6G) Internet of Things (IoT) wireless networks \cite{Lu}  emphasizes the importance of enhanced localization accuracy, which challenges us to innovate and rethink existing localization methods. In this context, reconfigurable intelligent surfaces (RIS) have emerged as a pivotal technology, significantly enhancing localization precision \cite{Cunhua1}. RIS offers an efficient and cost-effective alternative to conventional base stations (BS) \cite{Wu1}, providing enhanced connectivity, especially in the harsh environment. Additionally, the slim and adaptable feature of RIS is ideal for integration within urban infrastructure, meeting the demands of 6G networks \cite{Zhi3}. These advantages have sparked extensive contributions in the area of  RIS-aided localization.

RIS-aided localization algorithms are primarily divided into two categories: the \emph{two-step method} \cite{Shubo,Xing}  and the \emph{fingerprint-based method} \cite{WuFingerprint}. The \emph{two-step method} focuses on estimating parameters like angle of arrival (AoA) and time of arrival (ToA), using geometric relationships to determine location  \cite{huang2018deep}. On the other hand, the \emph{fingerprint-based method} \cite{3DCNN} operates by creating a database of pre-recorded signal characteristics, such as received signal strength (RSS), and then compares real-time signals with this database to identify a location. Building upon these two methods, research on algorithm design for RIS-aided localization has been studied extensively, exploring a range of applications and perspectives. { This  encompasses various specialized scenarios: multi-user localization employing deep learning techniques \cite{Kamran2, WuFingerprint}, enhanced localization performance under multi-RIS conditions \cite{Wu2RIS}, specific algorithms for near-field scenarios \cite{Shubo,Xing}, and the application of sidelink techniques and classical maximum likelihood methods in highway scenarios \cite{chen2023multirisenabled}.}

{Although RIS-aided localization algorithms have been extensively studied, the passive nature of RIS typically restricts these algorithms to being designed primarily based on the received signal at the BS, i.e., BS information. Nevertheless, the high cost and power consumption associated with radio-frequency chains often leads to a limited number of antennas at the BS, resulting in lower-dimensional BS information. This constraint can adversely affect the ability to accurately distinguish locations. }

{An important but overlooked aspect is that the received signal at the RIS, i.e., RIS information, actually contains much richer feature information about the user location than the BS signal, which can be exploited to achieve higher localization accuracy. This is because the RIS is commonly comprised of a large number of reflecting elements, and therefore, the dimension of its received signal could be high enough to store sufficient feature information, such as AoAs and ToAs, related to user locations. Utilizing this high-dimensional RIS information could unlock a new degree of freedom (DoF) in the algorithm design for RIS-aided localization, potentially leading to significantly higher localization accuracy. Recognizing this promising potential, our work focuses on leveraging high-dimensional RIS information for localization, which, to the best of our knowledge, has not been comprehensively addressed in the existing literature. }

Recognizing the promising potential, designing localization algorithms based on RIS information remains a challenging task. This complexity stems from the difficulty in reconstructing the high-dimensional RIS information. The passive nature of RIS makes it challenging to actively gather signals received at the RIS.  Furthermore, using low-dimensional BS information to derive high-dimensional RIS information results in non-unique outcomes, as each BS data can correspond to multiple distinct RIS information. The uncertainty in the signal reconstruction process requires innovative algorithms to accurately acquire the RIS information.

{To tackle this challenge, we employ the data-driven methods. i.e., convolution neural network (CNN), to design an RIS information reconstruction (RIS-IR) algorithm. Consequently, transfer learning is employed to design a fingerprint based algorithm to localize the mobile user (MU). The RIS-IR algorithm includes a data processing module for preprocessing low-dimensional BS information, a CNN module for feature extraction, and an output module to produce the reconstructed high-dimensional RIS information. We propose a transfer learning based fingerprint (TFBF) algorithm utilizing reconstructed high-dimensional RIS information for localization.  Specifically, we adapt a pre-trained DenseNet-121 model to map the reconstructed signal received at the RIS to the MU's   position. Simulation results reveal that the localization accuracy of the proposed TFBF algorithm is enhanced by utilizing the high-dimensional RIS information.
}
\section{System Model } \label{System_Model}
We consider an  RIS-aided uplink (UL) localization system where an  MU transmits the pilot signals to the  BS via the RIS. The BS is equipped with a uniform planar array (UPA) comprising $M_1\times M_2=M$ antennas, while the MU has a single antenna. Additionally, the RIS is equipped with a UPA comprising $N_1\times N_2=N$ reflecting elements.
\subsection{Geometry and Channel Model}
We assume that the BS is located at $\bm{p}_b = [x_b, y_b, z_b]^T$, while the center of the  RIS  is situated at  $\bm{p}_r = [x_r, y_r, z_r]^T$. Besides, it is assumed that the MU's actual position is $\bm{p}_u = [x_u, y_u, z_u]^T$. Typically, once the RIS and BS are installed, their positions, $\bm{p}_r$ and $\bm{p}_b$, are fixed and known.  To motivate the deployment of the RIS, we consider a common scenario where the line-of-sight (LoS) path between the BS and MU is obstructed \cite{Kangda1}. Such obstructions can result from various factors, including buildings, walls, or natural barriers \cite{Zhi4}\footnote{ {
System model described here can be straightforwardly applied to analog beamforming systems (e.g., Rice fading channel). Exploring the RIS information localization under the Rice fading channel would be an intriguing avenue for future work. }}.

 Next, let us model the MU-RIS and RIS-BS links. For any antenna array, given an elevation angle $\theta \in (0, \pi]$ and an azimuth angle $\phi \in (0, \pi]$, the array response vector can be typically described as
\begin{align}\label{gen_formula}
{\bm a}_{x}(\theta,\phi)={\bm a}_{x}^{(e)}(\theta)\otimes{\bm a}_{x}^{(a)}(\theta,\phi),
\end{align}
where $x$ indicates the specific link and $\otimes$ represents the Kronecker product. The components are defined as follows:
\begin{align}
{\bm a}_{x}^{(e)}(\theta) &= \left[1, \ldots, e^{\frac{-j2\pi(N_{e}-1) d_x\cos\theta}{\lambda_c}}\right]^T, \\
{\bm a}_{x}^{(a)}(\theta,\phi) &= \left[1, \ldots, e^{\frac{-j2\pi(N_{a}-1) d_x\sin\theta\cos\phi}{\lambda_c}}\right]^T,
\end{align}
where $N_{e}$ and $N_{a}$ denote the numbers of elements in the elevation and azimuth directions, respectively, $d_x$ denotes the distance between adjacent array elements, and $\lambda_c$ is the carrier wavelength.

\subsubsection{MU-RIS link}
Considering $P$ propagation paths between the MU and RIS, and utilizing the general array response vector defined in \eqref{gen_formula}, the channel response vector of the MU-RIS link can be expressed as
\begin{align}
{\bm g}_{ur} = \sum^{P}_{p=1}\alpha_{p}{\bm a}_{R_a}(\theta_{p},\phi_{p}),
\end{align}
where ${\bm a}_{R_a}$ denotes the array response vector of the RIS and $\alpha_{p}$ represents the channel gain of the $p$-th path.

\subsubsection{RIS-BS link}

Assuming $J$ propagation paths between the RIS and BS, the channel response matrix of the RIS-BS link can be modeled as
\begin{align}
{\bm H}_{rb}=\sum_{j=1}^{J}\beta_{j}{\bm a}_{B}(\theta_{j},\phi_{j}){\bm a}^{H}_{R_d}(\psi_{j},\omega_{j}),
\end{align}
where $\beta_{j}$ represents the channel gain of the $j$-th path, ${\bm a}_{B}$ and ${\bm a}^{H}_{R_d}$  are the array response vectors of the BS and RIS, respectively.

\subsection{Signal Model}
Letting ${\bm y}_r$ denote the RIS received signal, we have
\begin{align}\label{16r}
{\bm y}_r={\bm g}_{ur}{ s},
\end{align}
where ${s}$ denotes the pilot signal transmitted by the MU. Then,  defining the phase shift vector of the RIS as ${\bm \omega} = [\omega_1, \cdots, \omega_N]^T \in \mathbb{C}^{N\times1}$,   the UL signal received by the BS can be expressed as
\begin{align}\label{16}
{\bm y}={\bm H}_{rb}\textrm{diag}({\bm \omega}){\bm g}_{ur}{ s}+{\bm n},
\end{align}
where $\textrm{diag}({\bm \omega})$ and ${\bm n}$ denote the phase shift matrix of the RIS and the zero-mean additive white Gaussian noise, respectively.

\section{RIS Information Reconstruction (RIS-IR) Algorithm}
\begin{figure}[t]
\centering
\includegraphics[width=1\linewidth]{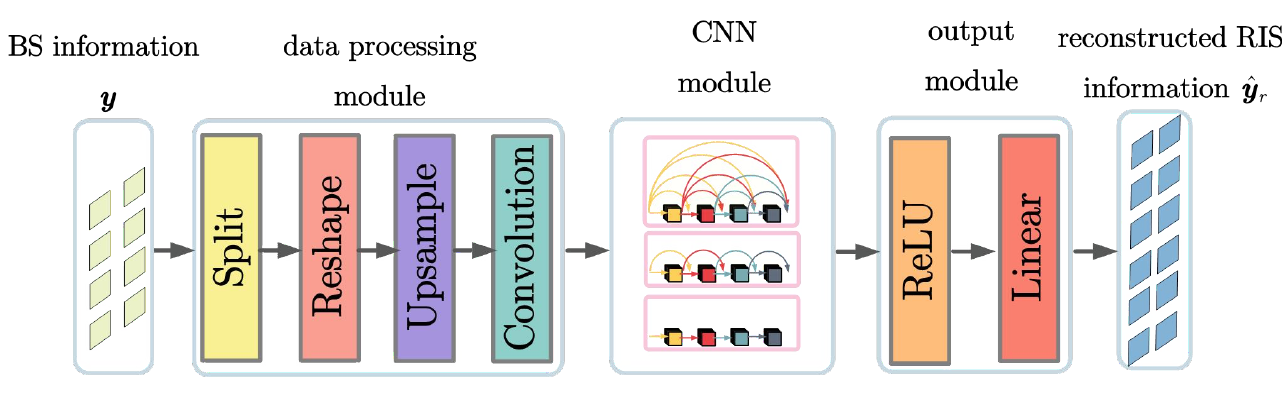}
\caption{\normalsize The structure of the RIS-IR algorithm.}\label{algorithm}
\end{figure}
 Traditionally, localization algorithm design is based on the BS information, e.g., ${\bm y}$. In this framework, we can define the estimated position of $\bm{{p}}_u$ as $\bm{\hat{p}}_u$, and then localization problem is  formulated as
  \begin{align}\label{prob}
  \underset{\bm{\hat{p}}_u }{\min} \ &|| \mathcal{H}( {\bm y})-\bm{{p}}_u ||^2.
\end{align}
The above problem can be solved with several localization algorithms which have been extensively investigated \cite{Shubo,Xing, WuFingerprint}.  

By defining the reconstructed high-dimensional RIS information as $\hat{\bm y}_r$, the RIS information reconstruction problem can be formulated as
\begin{equation} \label{pro1}
\min_{ \hat{\bm y}_r} \ \lVert {\bm y}_r - \hat{\bm y}_r\rVert^2_2,
\end{equation}
 which cannot be solved with traditional convex optimization methods, since $\hat{\bm y}_r$ is unknown.  Hence, we design a RIS information reconstruction (RIS-IR) algorithm to solve Problem \eqref{pro1}. As shown in Fig. \ref{algorithm}, the RIS-IR algorithm includes three modules: data processing module, CNN module, and output module, whose details are given in the following.
\subsection{Data Processing Module}
{The data processing module is designed to process the input BS information $\bm{y}$. Since $\bm{y}$ is a complex vector, we  divide it into its real and imaginary parts, represented as $\mathfrak{R}({\bm y})$ and $\mathfrak{I}( {\bm y} )$, respectively. Hence, we can convert  the complex-valued data into  real-valued input which is compatible with standard neural network architectures. Then,  since $\bm{y}$ consists of the features of the horizontal angle domain and the vertical angle domain of the BS, we reshape the two vectors $\mathfrak{R}({\bm y})\in \mathbb{C}^{M\times 1}$ and $\mathfrak{I}( {\bm y} )\in \mathbb{C}^{M\times 1}$ into matrices, denoted as ${\bm M}_{Ry}\in \mathbb{R}^{ M_1 \times M_2}$ and ${\bm M}_{Iy}\in \mathbb{R}^{ M_1 \times M_2}$, respectively. This reshaping aligns the data with the spatial dimensions of the BS.  Then, we concatenate the reshaped matrices of the real and imaginary parts,  preserving the phase and amplitude information necessary for accurate signal reconstruction. }Hence, the 3D tensor represented as
\begin{equation}\label{ty}
\boldsymbol{T}_{y} = \begin{bmatrix}
{\bm M}_{Ry}  \\
{\bm M}_{Iy}
\end{bmatrix}\in \mathbb{R}^{2 \times M_1 \times M_2}.
\end{equation}
However, the values of $M_1$ and $M_2$ may not be directly compatible with CNNs' architectures. Hence, an upsampling technique is applied:
\begin{equation} \label{tup}
\boldsymbol{T}_{\text{up}} = \text{Upsample}(\boldsymbol{T}_{y}) \in \mathbb{R}^{2 \times 256 \times 256}.
\end{equation}
{The upsampling technique is employed to adjust the dimensions of the tensor $\boldsymbol{T}_{y} $  to the desired dimensions of $2 \times 256 \times 256$. } It typically involves pixel value interpolation to expand the image or tensor size, redistributing pixel values accordingly. To make $\boldsymbol{T}_{\text{up}}$ suitable for processing by CNNs, which expects an input tensor of the form $(3,256,256)$, a convolutional (Con) layer is introduced to transform the bi-channel tensor to a tri-channel one:
\begin{equation}\label{tdense}
\boldsymbol{T}_{\text{Con}} = \text{Con}(\boldsymbol{T}_{\text{up}}).
\end{equation}
\subsection{CNN Module}
The CNN module is designed to extract the features of the input tensor $\boldsymbol{T}_{\text{Con}}$.  Drawing inspiration from super-resolution (SR) in the field of computer science (CS) \cite{super-resolution}, where high-resolution images are generated from their low-resolution counterparts using CNNs, we have adapted the  CNN architectures for feature extraction from low-dimensional matrices.
{This tensor format, corresponding to the real and imaginary parts of the BS signal and its UPA dimensions, aligns well with CNN's strengths in handling image-like data structures, making them ideal for extracting precise spatial features necessary for RIS information reconstruction.}

CNNs are renowned for their robust feature extraction capabilities, especially from low-resolution inputs. {The modularity of CNNs allows us to extend these basic models to more sophisticated architectures, facilitating future innovations in RIS-based localization systems.} This adaptability makes CNNs an essential tool in our study, aiming to leverage RIS information effectively for enhanced localization accuracy.
Several CNN architectures can be utilized to map low-dimensional data to high-dimensional representations, such as AlexNet \cite{super-resolution}, ResNet \cite{WuFingerprint}, and DenseNet \cite{densenet}. To further explore the potential of CNN for reconstructing RIS information,  we will compare their performance through simulation results. For simplicity, we refer to these architectures collectively as ``CNN" in our analysis.
\subsection{Output Module}
After the  CNN module, we design an output module to output the reconstructed RIS information. Within the output module, a ReLU layer is  used to introduce the non-linearity, enabling the network to capture and express the complex patterns of the high-dimensional RIS information. Besides,  the ReLU layer  can  address the vanishing gradient problem, facilitating better learning in deep neural networks.

Following the ReLU layer, a  {linear} layer is employed for directly outputting the desired high-dimensional RIS information. This layer serves to consolidate the complex features extracted by prior layers and transform them into a specific high-dimensional format, such as the required RIS information, effectively aligning the network's output with the target data structure. 

\subsection{Complexity Analysis}
{Data processing module involves reshaping and upsampling the input BS information. The reshaping operation has a linear complexity of \(O(M)\), where \(M\) is the dimension of the input vector \(\bm{y}\). The upsampling operation, which typically involves interpolation, has a complexity of \(O(N)\), where \(N\) is the size of the resulting upsampled tensor. Assuming the CNN module has \(L\) layers, each with \(K\) filters, and considering standard convolutional operations, the complexity can be approximated as \(O(L \cdot K \cdot F^2 \cdot C)\), where \(F\) is the filter size and \(C\) is the input channel size. The output module includes a ReLU layer  }

\section{RIS Information Based Localization Algorithm Design}
The RIS information contains sufficient feature information, e.g., AoAs and ToAs, related to user locations. Hence, in this section, we employ the fingerprint-based method \cite{WuFingerprint} to determine the MU's location using $\hat{\bm y}_r$, as obtained in the last section.
\subsection{Localization Problem Formulation}
Based on the fingerprint-based algorithm definition, the RIS information $\hat{\bm y}_r$, can be viewed as a unique fingerprint that possesses a specific mapping relationship with the MU. Hence,  we have
\begin{align}\label{26}
\bm{\hat{p}}_u=\mathcal{Q}(\hat{\bm y}_r )= [\hat{x}_u, \hat{y}_u, \hat{z}_u]^T,
\end{align}
where $\mathcal{Q}(\cdot)$ represents the  non-linear mapping function between $ \hat{\bm y}_r$ and  $\bm{\hat{p}}_u$.
Accordingly, the localization problem can be formulated as
\begin{align}\label{27b}
  \underset{\bm{\hat{p}}_u }{\min} \ &|| \mathcal{Q}(\hat{\bm y}_r  )-\bm{{p}}_u ||^2.
\end{align}
\subsection{Transfer Learning Based Fingerprint Algorithm}
To address Problem \eqref{27b} more efficiently with requiring fewer training epochs compared to traditional deep learning methods \cite{CNNsurvey}, we propose a  transfer learning based fingerprint (TFBF) algorithm to localize the MU \cite{transfer_learning}. By doing so, we can harness the knowledge of the model acquired from previous tasks, such as ImageNet.  Specifically, this algorithm employs the DenseNet-121 model, which has been pre-trained, to effectively map $\hat{\bm y}_r$ to the 3D position of the MU. We will first detail the structure and functionality of each component of the TFBF algorithm as follows.
\subsubsection{Input Module}
To extract features from $\hat{\bm y}_r$, we first separate it into its real and imaginary components, denoted as $\mathfrak{R}(\hat{\bm y}_r)$ and $\mathfrak{I}(\hat{\bm y}_r)$, respectively. These components are then combined and reshaped according to the UPA elements configuration at the RIS, forming a structure with dimensions $2\times N_1\times N_2$. Then, we employ upsampling techniques to adjust the signal dimensions, reshaping them to $2\times 256\times 256$ for compatibility with DenseNet. Finally, a Con layer is employed to transform this $2$ channel input into a $3$ channel format, making it suitable for DenseNet processing.
\subsubsection{Transferred DenseNet 121}
After preprocessing the received signal $\hat{\bm y}_r$, the  {input module} forwards the resulting tensor to the adapted DenseNet 121.  The choice of DenseNet 121 for this algorithm is motivated by its unique architecture, which draws inspiration from the principles of ResNet \cite{WuFingerprint}. Unlike ResNet's skip connections that may bypass several layers, DenseNet establishes direct connections between every layer, enhancing robust feature propagation. This interconnected design not only reduces the number of parameters, making the network more efficient, but also potentially allows DenseNet to outperform other deep neural networks, especially in capturing complex patterns in the data, which is essential for precise localization in RIS-aided systems.

Besides, it is worthy noted that the internal layers of the adapted DenseNet 121 are initially frozen in the initial stages of training and are progressively unfrozen as training advances.
\subsubsection{Output Module}
Following the transferred DenseNet 121, the output module is designed to estimate the location of the MU. Within the output module, the pooling layer is  utilized to reshape the output of the transferred DenseNet 121. Adding the pooling layer between the transferred DenseNet 121   and the linear layer avoids the large number of weight parameters introduced by the linear layer.  As a result, the AP layer reduces overfitting while improving the convergence rate.

After the pooling layer, a linear layer is employed to merge the features extracted from DenseNet 121. This layer consists of three neurons, each corresponding to the one coordinate of the MU's location.
\subsubsection{Training of TFBF Algorithm}
Leveraging the foundational weights and configurations from pre-trained models, the focus shifts to training the TFBF algorithm for the task of localization. Upon convergence of the training phase, the TFBF algorithm is ready to provide estimation for the MU's 3D position. For every reconstructed RIS information $\hat{\bm y}_r$, the network outputs a three-dimensional vector, $[\hat{x}_u, \hat{y}_u, \hat{z}_u]^T$. Therefore, Problem \eqref{27b} has been effectively solved by the proposed TFBF algorithm.

\section{Simulation Results}
In this simulation, we consider a setup in a 3D space, involving an MU, an RIS, and a BS. The carrier frequency is $f=90$ GHz. The RIS is composed of $ N_1 \times N_2 = 100$ elements, with adjacent elements spaced at $d_r=\lambda/2=1.67\times10^{-3}$ m. The center of the RIS is located at $\bm{p}_r=(15, 0, 2)$ m. The BS comprises $ M_1 \times M_2 = 9$ antennas, with an antenna spacing of $d_b=\lambda/2=1.67\times10^{-3}$ m. The center of the BS is positioned at $\bm{p}_b=(0, 10, 1.5) $ m.
The MU-RIS link includes $ P = 10 $ propagation paths, and the RIS-BS link comprises $ J = 10 $ propagation paths. The transmit power of the MU is $30$ dBm.

\begin{figure}[t]
    \centering
    \begin{minipage}{0.45\linewidth}
        \centering
        \includegraphics[width=\textwidth]{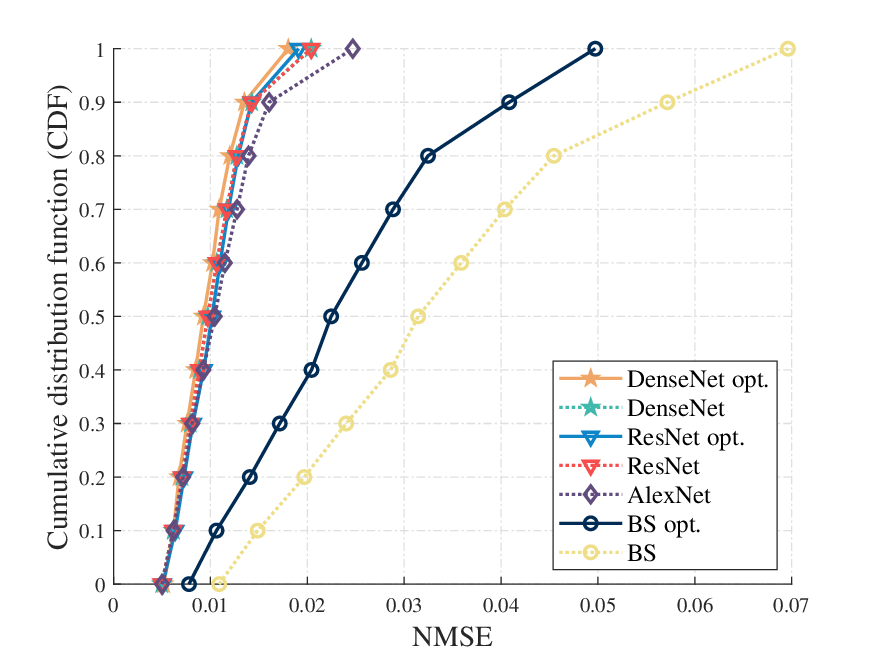}
        \caption{Localization accuracy comparison between using RIS information  and BS information.}
        \label{RIS}
    \end{minipage}
    \hfill
    \begin{minipage}{0.45\linewidth}
        \centering
        \includegraphics[width=\textwidth]{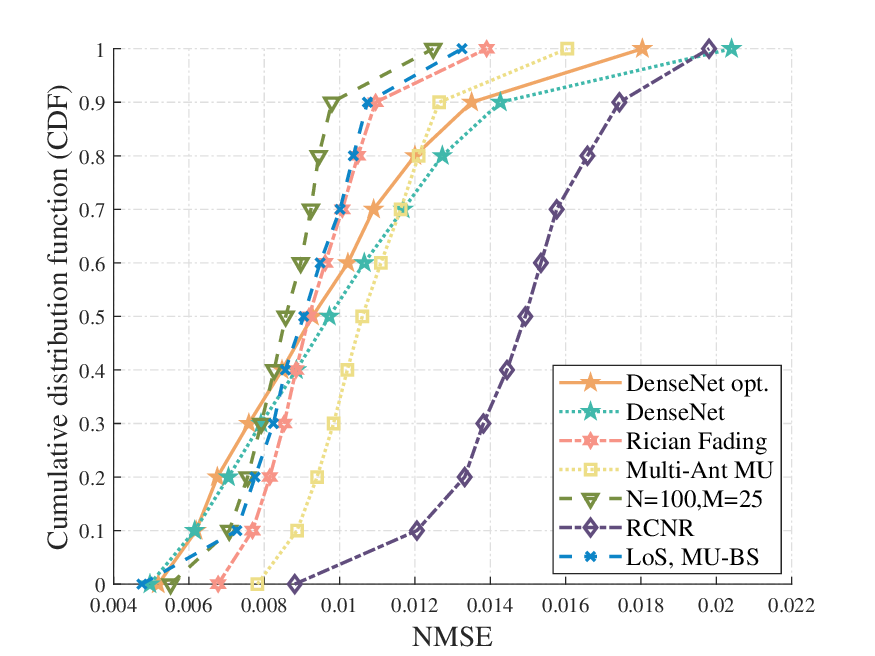}
        \caption{Localization accuracy comparison with different conditions.}
        \label{RISdif}
    \end{minipage} 
\end{figure}

In Fig. \ref{RIS}, we present  the normalized mean square error (NMSE)  cumulative distribution function (CDF) comparison to evaluate the localization accuracy achieved by various CNN architectures in reconstructing RIS information. These architectures include AlexNet, ResNet, and DenseNet \footnote{{ In the event of overfitting, dropout layers can be added to the CNN architecture as a regularization technique.   }}. As depicted in Fig. \ref{RIS}, { deeper CNN architectures, such as DenseNet, are more effective at reconstructing RIS information, thereby enhancing localization accuracy. However, the performance improvement of DenseNet, despite being a deeper network, is only marginally better than ResNet. \textbf{\emph{This observation indicates a need for future research to identify more optimal neural network architectures that can more effectively decode RIS information for improved localization.}} Additionally, this finding highlights the efficiency of even relatively shallower networks like AlexNet in handling RIS information, suggesting a balance between model complexity and performance.}

Besides, we also provide the  localization accuracy comparison between using  BS information and utilizing RIS information for localization. In Fig. \ref{RIS}, the CDF of NMSE using BS information is denoted as `BS'. As shown in the figure, the NMSE for `BS'   at  $90\%$ is $0.058$,  indicating that relying on BS information for localization is significantly less accurate than using RIS information. {This error is substantially higher than the NMSE under `AlexNet', which represents  the least effective CNN architecture for reconstructing RIS information.
\emph{\textbf{It clearly demonstrates the substantial benefits of employing high-dimensional RIS information in RIS-aided localization.}} Consequently, \textbf{\emph{this demonstrates the concept that RIS information opens new DoF in the algorithm design of RIS-aided localization systems.} } The superior performance of RIS information underscores its critical role in enhancing localization accuracy and robustness, especially when compared to traditional BS-based localization methods.}

Furthermore, Fig. \ref{RIS} presents curves illustrating scenarios where the phase shift matrix of the RIS is optimized to maximize the received SNR at the  BS \cite{Wuqq1}. In these simulations, the curve labeled `BS opt.' directly represents the scenarios with the highest received SNR at the BS. Additionally, the curves labeled  `DenseNet opt.' and  `ResNet opt.' are generated from localization results using RIS information, which has been reconstructed by employing the DenseNet-121 and ResNet-18 models, respectively.
When comparing these optimized scenarios, we observe only a marginal difference. However, a significant gap is noticed between `BS opt.' and the standard `BS', highlighting the significant improvement in localization accuracy with optimized phase shifts when relying solely on BS information. This observation emphasizes the robustness of RIS information against unoptimized phase shifts.  \textbf{\emph{This suggests that the high-dimensional RIS information predominantly determines localization accuracy, making the phase shift design less critical in this context. }}

{To validate the generalization capabilities of our proposed algorithm, we conducted several comparative analyses under different simulation settings as depicted in Fig.~\ref{RISdif}. Initially, to assess the algorithm's suitability for analog beamforming-based systems, we implemented a Rician fading channel, which supports the algorithm well with a performance enhancement. Additionally, two scenarios were tested: first, we assumed that the MU is equipped with two antennas, and second, a LoS link was added from the MU to the BS. Both modifications showed improvement in localization accuracy. Furthermore, we expanded the BS's antenna array to \(M = 25\). This adjustment significantly improved the results, likely due to the increased information available for reconstructing more accurate RIS information, thus enhancing localization precision. Finally, we compared our proposed TFBF algorithm with the RCNR localization algorithm presented in \cite{WuFingerprint}, demonstrating that our TFBF algorithm achieves higher precision.
}
\section{Conclusion}\label{Con}
 We have proposed an RIS-IR  algorithm to reconstruct the RIS information with BS information.  We have also proposed a   TFBF algorithm that employed reconstructed RIS information for MU localization. The TFBF algorithm involved adapting a pre-trained DenseNet-121 model to map the reconstructed RIS signal to the MU's 3D position. Empirical results have demonstrated that the performance of the localization algorithm is dominated by the high-dimensional RIS information and is robust to unoptimized phase shifts.

\bibliographystyle{IEEEtran}
				\bibliography{myre}

\begin{thebibliography}{10}
\providecommand{\url}[1]{#1}
\csname url@samestyle\endcsname
\providecommand{\newblock}{\relax}
\providecommand{\bibinfo}[2]{#2}
\providecommand{\BIBentrySTDinterwordspacing}{\spaceskip=0pt\relax}
\providecommand{\BIBentryALTinterwordstretchfactor}{4}
\providecommand{\BIBentryALTinterwordspacing}{\spaceskip=\fontdimen2\font plus
\BIBentryALTinterwordstretchfactor\fontdimen3\font minus
  \fontdimen4\font\relax}
\providecommand{\BIBforeignlanguage}[2]{{%
\expandafter\ifx\csname l@#1\endcsname\relax
\typeout{** WARNING: IEEEtran.bst: No hyphenation pattern has been}%
\typeout{** loaded for the language `#1'. Using the pattern for}%
\typeout{** the default language instead.}%
\else
\language=\csname l@#1\endcsname
\fi
#2}}
\providecommand{\BIBdecl}{\relax}
\BIBdecl

\bibitem{Lu}
S.~Lu \emph{et~al.}, ``{Integrated Sensing and Communications: Recent Advances
  and Ten Open Challenges},'' \emph{IEEE Internet Things J.}, pp. 1--1, 2024.

\bibitem{Cunhua1}
C.~Pan \emph{et~al.}, ``{An Overview of Signal Processing Techniques for
  RIS/IRS-Aided Wireless Systems},'' \emph{IEEE J. Sel. Topics Signal
  Process.}, vol.~16, no.~5, pp. 883--917, 2022.

\bibitem{Wu1}
T.~Wu \emph{et~al.}, ``{Joint Angle Estimation Error Analysis and 3D
  Positioning Algorithm Design for mmWave Positioning System},'' \emph{IEEE
  Internet Things J.}, pp. 1--1, 2023.

\bibitem{Zhi3}
K.~Zhi \emph{et~al.}, ``{Two-Timescale Design for Reconfigurable Intelligent
  Surface-Aided Massive MIMO Systems With Imperfect CSI},'' \emph{IEEE Trans.
  Inf. Theory}, vol.~69, no.~5, pp. 3001--3033, 2023.

\bibitem{Shubo}
S.~Huang \emph{et~al.}, ``{Near-Field RSS-Based Localization Algorithms Using
  Reconfigurable Intelligent Surface},'' \emph{IEEE Sens. J.}, vol.~22, no.~4,
  pp. 3493--3505, 2022.

\bibitem{Xing}
X.~Zhang \emph{et~al.}, ``{Hybrid Reconfigurable Intelligent Surfaces-Assisted
  Near-Field Localization},'' \emph{IEEE Commun. Lett.}, vol.~27, no.~1, pp.
  135--139, 2023.

\bibitem{WuFingerprint}
T.~Wu \emph{et~al.}, ``{Fingerprint-Based mmWave Positioning System Aided by
  Reconfigurable Intelligent Surface},'' \emph{IEEE Wirel. Commun. Lett.},
  vol.~12, no.~8, pp. 1379--1383, 2023.

\bibitem{huang2018deep}
H.~Huang \emph{et~al.}, ``{Deep Learning for Super-Resolution Channel
  Estimation and DOA Estimation Based Massive MIMO System},'' \emph{IEEE Trans.
  Veh. Technol.}, vol.~67, no.~9, pp. 8549--8560, 2018.

\bibitem{3DCNN}
C.~Wu, X.~Yi, W.~Wang, L.~You, Q.~Huang, X.~Gao, and Q.~Liu, ``{Learning to
  Localize: A 3D CNN Approach to User Positioning in Massive MIMO-OFDM
  Systems},'' \emph{IEEE Trans. Wireless Commun.}, vol.~20, no.~7, pp.
  4556--4570, 2021.

\bibitem{Kamran2}
K.~Keykhosravi \emph{et~al.}, ``{RIS-Enabled SISO Localization Under User
  Mobility and Spatial-Wideband Effects},'' \emph{IEEE J. Sel. Topics Signal
  Process.}, vol.~16, no.~5, pp. 1125--1140, 2022.

\bibitem{Wu2RIS}
T.~Wu \emph{et~al.}, ``{3D Positioning Algorithm Design for RIS-aided mmWave
  Systems},'' 2022.

\bibitem{chen2023multirisenabled}
H.~Chen, P.~Zheng, M.~F. Keskin, T.~Al-Naffouri, and H.~Wymeersch,
  ``{Multi-RIS-Enabled 3D Sidelink Positioning},'' 2023.

\bibitem{Kangda1}
K.~Zhi, C.~Pan, H.~Ren, K.~K. Chai, and M.~Elkashlan, ``{Active RIS Versus
  Passive RIS: Which is Superior With the Same Power Budget?}'' \emph{IEEE
  Commun. Lett.}, vol.~26, no.~5, pp. 1150--1154, 2022.

\bibitem{Zhi4}
K.~Zhi \emph{et~al.}, ``{Power Scaling Law Analysis and Phase Shift
  Optimization of RIS-Aided Massive MIMO Systems With Statistical CSI},''
  \emph{IEEE Trans. on Commun.}, vol.~70, no.~5, pp. 3558--3574, 2022.

\bibitem{super-resolution}
S.~Farsiu, D.~Robinson, M.~Elad, and P.~Milanfar, ``{Advances and challenges in
  super-resolution},'' \emph{Int. J. Imag. Syst. Technol.}, vol.~14, no.~2, pp.
  47--57, 2004.

\bibitem{densenet}
G.~Huang, Z.~Liu, L.~Van Der~Maaten, and K.~Q. Weinberger, ``{Densely Connected
  Convolutional Networks},'' in \emph{Proc. IEEE Conf. Comput. Vis. Pattern
  Recognit.}, 2017, pp. 2261--2269.

\bibitem{CNNsurvey}
Z.~Li \emph{et~al.}, ``{A Survey of Convolutional Neural Networks: Analysis,
  Applications, and Prospects},'' \emph{IEEE Trans. Neural Netw. Learn. Syst.},
  vol.~33, no.~12, pp. 6999--7019, 2022.

\bibitem{transfer_learning}
S.~J. Pan and Q.~Yang, ``{A Survey on Transfer Learning},'' \emph{IEEE Trans.
  Knowl. Data Eng.}, vol.~22, no.~10, pp. 1345--1359, 2010.

\bibitem{Wuqq1}
Q.~Wu and R.~Zhang, ``{Intelligent Reflecting Surface Enhanced Wireless Network
  via Joint Active and Passive Beamforming},'' \emph{IEEE Trans. Wireless
  Commun.}, vol.~18, no.~11, pp. 5394--5409, 2019.

\end{thebibliography}

			\end{document}